\documentclass[letterpaper, 10 pt, conference]{ieeeconf}  

\IEEEoverridecommandlockouts                              

\usepackage{booktabs}
\usepackage{makecell}
\usepackage{multirow}
\usepackage{color}
\newcommand{\greencheckmark}{\textcolor{green}{\checkmark}}
\newcommand{\redxmark}{\textcolor{red}{\times}}
\usepackage{arydshln}

\usepackage{graphics} 
\usepackage{epsfig} 
\usepackage{mathptmx} 
\usepackage{times} 
\usepackage{amsmath} 
\usepackage{amssymb}  

\title{\LARGE \bf
Data-Free Distillation Improves Efficiency and Privacy in Federated Thorax Disease Analysis
}

\author{Ming Li$^{1,2}$ and Guang Yang$^{1,2,3,4}$ 
\thanks{
$^{1}$Bioengineering Department and Imperial-X, Imperial College London, London W12 7SL, UK.
$^{2}$National Heart and Lung Institute, Imperial College London, London SW7 2AZ, UK.
$^{3}$Cardiovascular Research Centre, Royal Brompton Hospital, London SW3 6NP, UK.
$^{4}$School of Biomedical Engineering \& Imaging Sciences, King's College London, London WC2R 2LS, UK. Corresponding Authors: Ming Li (ming.li@imperial.ac.uk) and Guang Yang (g.yang@imperial.ac.uk)}%
\thanks{This study was supported in part by the ERC IMI (101005122), the H2020 (952172), the MRC (MC/PC/21013), the Royal Society (IEC\textbackslash NSFC\textbackslash211235), the NVIDIA Academic Hardware Grant Program, the SABER project supported by Boehringer Ingelheim Ltd, Wellcome Leap Dynamic Resilience, and the UKRI Future Leaders Fellowship (MR/V023799/1).}
\thanks{This study utilized the publicly available NIH Chest X-ray 14 dataset. Ethical approval for the collection and distribution of this dataset was previously obtained by the originating institution.}%
}

\begin{document}

\maketitle
\thispagestyle{empty}
\pagestyle{empty}

\begin{abstract}
Thorax disease analysis in large-scale, multi-centre, and multi-scanner settings is often limited by strict privacy policies. 
Federated learning (FL) offers a potential solution, while traditional parameter-based FL can be limited by issues such as high communication costs, data leakage, and heterogeneity. 
Distillation-based FL can improve efficiency, but it relies on a proxy dataset, which is often impractical in clinical practice.
To address these challenges, we introduce a data-free distillation-based FL approach FedKDF. 
In FedKDF, the server employs a lightweight generator to aggregate knowledge from different clients without requiring access to their private data or a proxy dataset. 
FedKDF combines the predictors from clients into a single, unified predictor, which is further optimized using the learned knowledge in the lightweight generator.
Our empirical experiments demonstrate that FedKDF offers a robust solution for efficient, privacy-preserving federated thorax disease analysis.
\newline

\indent \textit{Clinical relevance}— This study offers a privacy-preserving method for multi-centre thorax disease analysis, improving diagnostics without compromising patient data.
\end{abstract}

\section{INTRODUCTION}

Medical data is often fragmented across multiple centres, compromising clinical insights, particularly for rare diseases [1,2]. Stringent privacy regulations further restrict large-scale, multi-centre, and multi-scanner analysis of thorax disease. Numerous methods have been proposed to address the privacy challenge, among which Federated Learning (FL) stands out as a promising solution by enabling cross-centre analysis without data transfer beyond individual firewalls.

Common FL algorithms like Federated Averaging (FedAvg) [3], which are based on parameter-averaging schemes, have certain limitations including the necessity for uniform model architectures across clients, high communication costs, and compromised performance due to non-iid (independent and identically distributed) data distributions. Distillation-based FL [4] mitigates some of these issues but relies on impractical proxy datasets in clinical settings.

To address the aforementioned challenges, we propose FedKDF, a data-free distillation FL approach that eliminates the need for a proxy dataset. 
In this method, clients upload their predictors to the central server for aggregation. 
The server then utilizes a lightweight conditional generator to create latent feature representations, which are consistent with the ensemble of client predictions.
Then the aggregated predictor is further optimized using the learned knowledge in the lightweight conditional generator.
The updated predictor is subsequently sent back to all clients for the next communication round. 
FedKDF combines the benefits of distillation-based FL without the constraints of a proxy dataset, making it a more practical and efficient option for large-scale, multi-centre, and multi-scanner thorax disease analysis.
Table I provides a comprehensive comparison of various non-FL and FL methods, along with their unique pros and cons.


\begin{table*}
\begin{center}
    \caption{Comparison of Non-FL and various FL settings.}
    \label{fl_approaches}
    \resizebox{0.77\textwidth}{!}{%
    \begin{tabular}{ccccccc}
    \toprule
    \multirow{2}{*}{Method} & \multicolumn{2}{c}{Information Exchanged} & \multirow{2}{*}{\makecell[c]{Proxy\\Dataset}} & \multirow{2}{*}{Privacy} & \multirow{2}{*}{\makecell[c]{Communication\\Efficiency}} & \multirow{2}{*}{\makecell[c]{Model\\Heterogeneity}} \\
    \cmidrule{2-3}

    & Upload & Download & & & \\

    \midrule
    Standalone & - & - & - & $\greencheckmark$ & - & - \\
    Centralized & - & - & - & $\redxmark$ & - & - \\
    \hdashline
    Parameter-based FL (FedAvg) & model parameters & model parameters & - & $\redxmark$ & $\redxmark$ & $\redxmark$ \\
    Distillation-based FL via proxy dataset (FedKD)  & logit vectors & logit vectors & $\greencheckmark$ & $\greencheckmark$ & $\greencheckmark$ & $\greencheckmark$ \\
    \makecell[c]{Distillation-based FL via data-free (FedKDF)} & partial model parameters & partial model parameters & - & $\greencheckmark$ & $\greencheckmark$ & $\greencheckmark$ \\
    \bottomrule
    \end{tabular}%
    }
\end{center}
\end{table*}

\section{METHOD}

In a typical FL setting, we have a set of clients $C$.
Each client $k$ has a local private dataset $D^{k} = \{(x_{i}, y_{i})\}^{n_{k}}_{i=1}$, where $n_{k} = |D^{k}|$ is the number of data samples belonging to user $k$.
Let $X \subset \mathbb{R}^{D}$ be the data space, $Z \subset \mathbb{R}^{d}$ be the latent feature space, and $Y \subset \mathbb{R}$ be the output space, where $d \ll D$.
The model is parameterized by $\theta = [\theta^{f}, \theta^{p}]$, with feature extractor $F_{\theta^{f}}(\cdot)$ and predictor $P_{\theta^{p}}(\cdot)$.
Federated Learning aims to learn a global model parameterized by $\theta$ that minimizes its risk on each of the clients:
\begin{equation}
    \min _{\boldsymbol{\theta}} \frac{1}{K} \sum_{k=1}^K L\left(P(F(x_{i};\theta^{f});\theta^{p})\right)
\end{equation}
where $L$ is the loss function.
\par

Typical knowledge distillation-based FL (FedKD) employs a proxy dataset $D^{p}$ to minimize the discrepancy between the logits outputs from the client models $\theta_{k}$ (teachers) and the global model $\theta$ (student).
Kullback-Leibler divergence $D_{KL}(\cdot)$ is usually used to measure such discrepancy:
\begin{equation}
    \resizebox{0.45\textwidth}{!}{$
        \underset{\boldsymbol{\theta}}{min} \underset{x \sim D^{P}}{\mathbb{E}} \left[ D_{KL} \left[ \sigma \left(\frac{1}{K} \sum_{k=1}^K g(F(x;\theta^{f}_{k});\theta^{p}_{k})\right) \parallel \sigma \left(g(F(x;\theta^{f});\theta^{p}) \right) \right] \right]
    $}
\end{equation}
where $g(\cdot)$ is the logits output of the predictor $P$, and $\sigma(\cdot)$ is the non-linear activation, and $P(z; \theta^{p}) = \sigma(g(z; \theta^{p}))$.
\par

To get rid of the proxy dataset, our FedKDF conducts knowledge distillation by learning a conditional generator $G$ parameterized by $\omega$:
\begin{equation}
    \resizebox{0.4\textwidth}{!}{$
        \underset{\boldsymbol{\omega}}{min} 
        \underset{y \sim p(y)}{\mathbb{E}} 
        \underset{z \sim G(z|y)}{\mathbb{E}} 
        \left[
            L \left(
                \sigma (\frac{1}{K} \sum_{k=1}^K g(z;\theta^{p}_{k}));y
            \right)
        \right]
    $}
\end{equation}
to recover the distribution over the latent feature space $Z$.
$p(y)$ is the ground-truth prior distributions of the target labels.
The learned $G$ further optimized the aggerated predictor $P$, which is then distributed to all clients for the next communication round.

\begin{figure}
    \centering
    \includegraphics[width=0.21\textwidth]{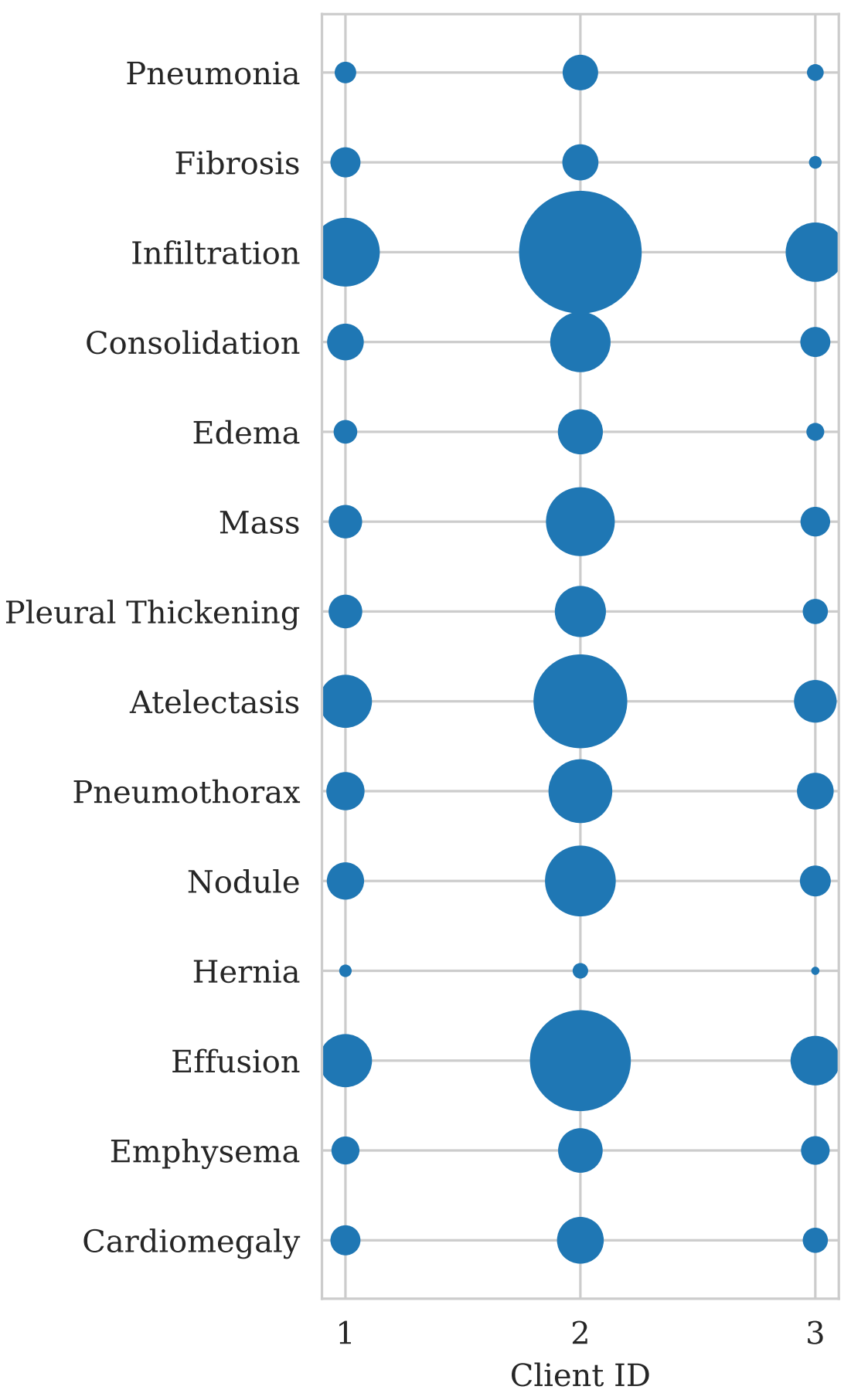}
    \caption{Visualization of data distribution per class allocated to each client.}
\end{figure}

\begin{table}
\centering
\resizebox{0.43\textwidth}{!}{%
\begin{tabular}{cccccc}
\toprule
\multirow{2}{*}{Pathology} & \multicolumn{5}{c}{Method} \\
\cmidrule{2-6}
& Centralized & Standalone & FedAvg & FedKD & FedKDF \\
\midrule
Atelectasis & 
77.89$\pm$0.98 & 67.83$\pm$1.17 & 73.64$\pm$1.21 & 75.48$\pm$1.19 & 74.97$\pm$1.22 \\

Cardiomegaly & 
89.58$\pm$1.21 & 75.12$\pm$2.15 & 81.23$\pm$2.10 & 86.87$\pm$1.52 & 86.19$\pm$1.57 \\

Consolidation & 
79.31$\pm$1.30 & 67.07$\pm$1.67 & 71.94$\pm$1.41 & 77.52$\pm$1.29 & 76.71$\pm$1.33 \\

Edema & 
87.54$\pm$1.45 & 67.40$\pm$3.02 & 76.05$\pm$2.87 & 84.57$\pm$1.48 & 83.82$\pm$1.52 \\

Effusion & 
86.81$\pm$0.67 & 70.30$\pm$1.00 & 77.92$\pm$0.94 & 83.72$\pm$0.76 & 82.98$\pm$0.81 \\

Emphysema & 
88.80$\pm$1.56 & 65.09$\pm$2.51 & 76.68$\pm$2.37 & 85.97$\pm$1.59 & 85.34$\pm$1.64 \\

Fibrosis & 
78.75$\pm$2.31 & 63.10$\pm$3.06 & 69.18$\pm$2.92 & 75.99$\pm$2.22 & 75.11$\pm$2.27 \\

Hernia & 
87.85$\pm$6.71 & 63.56$\pm$8.96 & 75.99$\pm$7.91 & 85.21$\pm$6.53 & 84.12$\pm$6.63 \\

Infiltration & 
67.70$\pm$0.95 & 58.15$\pm$0.98 & 62.97$\pm$0.92 & 65.79$\pm$0.91 & 65.21$\pm$0.95 \\

Mass & 
83.02$\pm$1.31 & 63.97$\pm$1.67 & 71.93$\pm$1.56 & 79.98$\pm$1.31 & 79.11$\pm$1.36 \\

Nodule & 
74.84$\pm$1.44 & 58.33$\pm$1.60 & 66.02$\pm$1.52 & 72.47$\pm$1.42 & 71.71$\pm$1.46 \\

Pleural Thickening & 
74.63$\pm$1.84 & 59.61$\pm$2.24 & 66.49$\pm$2.11 & 72.03$\pm$1.81 & 71.24$\pm$1.86 \\

Pneumonia & 
70.06$\pm$3.20 & 46.96$\pm$3.70 & 58.03$\pm$3.51 & 67.04$\pm$3.12 & 66.09$\pm$3.18 \\

Pneumothorax & 
84.10$\pm$1.30 & 64.80$\pm$1.75 & 75.52$\pm$1.62 & 82.01$\pm$1.34 & 81.09$\pm$1.38 \\

\midrule
mAUC & 
84.83$\pm$0.28 & 72.16$\pm$0.40 & 74.01$\pm$0.39 & 82.71$\pm$0.29 & 81.92$\pm$0.33 \\
\bottomrule

\end{tabular}%
}
\caption{Comparison results (AUC $\pm$ $95\%$ CI).}
\label{tab:comparison_sotas}
\end{table}

\begin{table}
\centering
\resizebox{0.15\textwidth}{!}{%
\begin{tabular}{|c|c|}
\hline
Method & Bandwidth (MB) \\
\hline
FedAvg & 86800.00 \\
FedKD  & 89.60 \\
FedKDF & 40.00 \\
\hline
\end{tabular}%
}
\caption{Communication cost to convergence.}
\label{table:comm_cost}
\end{table}

\section{EXPERIMENTS}
We conducted our experiments using the NIH Chest X-ray 14 dataset [5], which consists of 112,120 X-ray images from 30,805 unique patients. The dataset encompasses 15 classes, including 14 diseases and one ``no findings" category.
To emulate client heterogeneity and non-iid conditions, we partitioned the training dataset into three client subsets using the Dirichlet distribution with a default $\alpha$ value of 1.0. The data distribution for each client, represented by dot size, is visualized in Figure 1.
The classification performance was evaluated using the Area Under the Curve (AUC) of the Receiver Operating Characteristic (ROC) curve.

We compared FedKDF against centralized training, standalone training, vanilla FedAvg, and FedKD methods. The FedKD method employed the Progressive Growing GAN (PGGAN) [6] to generate a synthetic proxy dataset.
Quantitative results are summarized in Table~\ref{tab:comparison_sotas}. Our proposed FedKDF achieves a mean AUC (mAUC) of 81.92$\pm$0.33\%, which is nearly as effective as FedKD (82.71$\pm$0.29\% mAUC) and even comparable with centralized training (84.83$\pm$0.28\% mAUC). The performance suggests that FedKDF offers a viable, privacy-preserving, and efficient solution.
We also assessed the communication cost for each method. Distillation-based FL methods, both FedKD and FedKDF, significantly reduce the communication burden. Specifically, FedKD required only 89.60 MB of data exchange for convergence, a significant drop from the 86800.00 MB required by vanilla FedAvg. While our proposed FedKDF outperformed all, needing just 40.00 MB for convergence. The detailed figures are provided in Table III.

\section{CONCLUSION}
This study offers a privacy-preserving method FedKDF for multi-centre thorax disease analysis, improving
diagnostics without compromising patient data.
The experimental results emphasize the efficacy of FedKDF in offering not just privacy preservation and robust performance, but also efficiency in communication costs.

\addtolength{\textheight}{-12cm}   






\begin{thebibliography}{99}

\bibitem{c1} Newton, Katherine M., et al. ``Validation of electronic medical record-based phenotyping algorithms: results and lessons learned from the eMERGE network." Journal of the American Medical Informatics Association 20.e1 (2013): e147-e154.
\bibitem{c3} McMahan, Brendan, et al. ``Communication-efficient learning of deep networks from decentralized data." Artificial intelligence and statistics. PMLR, 2017.
\bibitem{c4} Lin, Tao, et al. ``Ensemble distillation for robust model fusion in federated learning." Advances in Neural Information Processing Systems 33 (2020): 2351-2363.
\bibitem{c5} Wang, Xiaosong, et al. ``Chestx-ray8: Hospital-scale chest x-ray database and benchmarks on weakly-supervised classification and localization of common thorax diseases." Proceedings of the IEEE conference on computer vision and pattern recognition, 2017.
\bibitem{c6} Karras, Tero, et al. ``Progressive growing of gans for improved quality, stability, and variation." International Conference on Learning Representations, 2018.




\end{thebibliography}
\end{document}